\documentclass[%
 reprint,
 superscriptaddress,
 showpacs,preprintnumbers,
 amsmath,amssymb,
 aps,
pra,
 floatfix,
]{revtex4-1}

\usepackage[T1]{fontenc} 
\usepackage{mathptmx}
\usepackage{graphicx}
\usepackage{dcolumn}
\usepackage{bm}
\usepackage[colorlinks,breaklinks,
 citecolor=red,linkcolor=red,urlcolor=blue
]{hyperref}

\begin{document}

\title{Optical stimulated slowing of polar heavy-atom molecules with a constant beat phase}%

\author{Yanning Yin}
\author{Supeng Xu}
\author{Meng Xia}
\affiliation{State Key Laboratory of Precision Spectroscopy, School of Physics and Materials Science, East China Normal University, Shanghai 200062, China}
\author{Yong Xia}
\email[]{yxia@phy.ecnu.edu.cn}
\affiliation{State Key Laboratory of Precision Spectroscopy, School of Physics and Materials Science, East China Normal University, Shanghai 200062, China}
\affiliation{NYU-ECNU Institute of Physics at NYU Shanghai, Shanghai 200062, China}
\author{Jianping Yin}
\affiliation{State Key Laboratory of Precision Spectroscopy, School of Physics and Materials Science, East China Normal University, Shanghai 200062, China}


\date{\today}

\begin{abstract}
Polar heavy-atom molecules have been well recognized as promising candidates for precision measurements and tests of fundamental physics. A much slower molecular beam to increase the interaction time should lead to a more sensitive measurement. Here we theoretically demonstrate the possibility of the stimulated longitudinal slowing of heavy-atom molecules by the coherent optical bichromatic force with a constant beat phase. Taking the YbF meolecule as an example, we show that a rapid and short-distance deceleration of heavy molecules by a phase-compensation method is feasible with moderate conditions. A molecular beam of YbF with a forward velocity of 120 m/s can be decelerated below 10 m/s within a distance of 3.5 cm and with a laser irradiance for each traveling wave of 107.2 W/cm$^2$. We also give a simple approach to estimate the performance of the BCF on some other heavy molecules, which is helpful for making a rapid evaluation on the feasibility of the stimulated slowing experiment. Our proposed slowing method could be a promising approach to break through the space constraint or the limited capture efficiency of molecules loadable into a MOT in traditional deceleration schemes, opening the possibility for a significant improvement of the precision measurement sensitivity.
\end{abstract}
\maketitle


\section{\label{section1}INTRODUCTION}

Cold or ultracold molecules are playing an increasingly important role in diverse research areas of modern physics\cite{Doyle2004,Carr2009,Jin2012,DeMille2017}, including quantum simulation\cite{Micheli2006} and information processing\cite{DeMille2002,Yelin2006}, studies of ultracold collisions and controlled chemistry\cite{Sawyer2011,Hummon2011,Krems2008}, precision measurements and tests of fundamental physics\cite{Hunter2012,Tarbutt2013}. Among these significant applications, polar diatomics containing heavy elements are well recognized as quite promising objects for the search of a break of inversion symmetry (P) and time-reversal invariance (T), which would indicate the presence of new physics beyond the Standard Model (SM) of particle physics that is of fundamental importance\cite{Erler2005}. For instance, measurements of the electron's electric dipole moment (eEDM) have been widely acknowledged as a powerful tool in this search for physics beyond the SM\cite{Fortson2003}, since the SM and theories that extend the SM predict quite different limits of the values of eEDM that should be detectable using current experimental techniques\cite{Pospelov1991}. Compared to atoms, polar heavy-atom molecules are generally a more advantageous platform for the eEDM measurements because of their stronger degree of polarization, which would not only considerably increase the interaction strength with the eEDM due to the internal electric field of a polar molecule\cite{Hinds1997} but also suppress important systematic errors arising from motional magnetic fields and geometric phases\cite{Hudson2002}. Therefore, a new generation of modern eEDM experiments employing polar molecules has been implemented during the past decade and has provided tighter constraints on the parameters of theories that extend the SM\cite{Baron2013}. The first molecular determination of the eEDM with YbF molecules obtained an upper limit on the eEDM of order of $10^{-28}\ e\ \rm{cm}$\cite{Hudson2011}, and the subsequent exploration with ThO molecules has improved it to $\left|d_e\right|<8.7\times10^{-29}\ e\ \rm{cm}$\cite{Baron2013},  which was then confirmed by the experimental demonstration with trapped HfF$^+$ ions\cite{Cairncross2017}. Some other candidates for the EDM experiments, such as HgF\cite{Prasannaa2015}, PbF\cite{ShaferRay2006}, PbO\cite{Eckel2013}, TlF\cite{Hunter2012}, have also been discussed or being prepared in recent years for a further reduction of the current limit, necessitating higher sensitivity and more stringent rejection of systematic errors.

A more precise measurement of eEDM by minimizing the statistical uncertainty in $d_e$ requires a more favorable internal effective electric field, $\rm{\bm{E}_{eff}}$, in the molecule, more intense sources of molecules with better state preparation, and a further increase in the interaction time, $\tau$, of molecules with the applied fields, as is indicated in Ref. \cite{Tarbutt2013}. Improvement of $\tau$ could be achieved by either extending the length of the interaction region, or introducing a much slower and colder molecular beam than the currently used molecular beams produced by supersonic expansion or buffer gas cooling at T $\approx$ 4 K which limited the $\tau\approx1$ ms\cite{Hudson2011,Baron2013}. Increasing the interaction length could lead to the considerable loss of the number of detectable molecules due to the divergence of the molecular beam, which could be effectively addressed with transverse laser cooling. Recently, transverse Doppler and sub-Doppler cooling of YbF molecules have been demonstrated by J. Lim \emph{et al} and a feasible coherence time exceeding 150 ms was obtained by cooling the YbF beam below 100 $\mu$K\cite{Lim2017}. However, they also pointed out that the current molecular beam from a cryogenic buffer gas source with a forward speed of $\sim$160 m/s would still require a 24 m-long experiment to make use of this ultracold molecules. In consideration of the space constraint, it is worthwhile to explore a practical and efficient method of longitudinally decelerating the heavy molecules produced from an available beam source to a quite low speed, and it would be of more significance to slow the molecular beam to the capture velocity of a 3D magneto-optical trap (MOT). By further reducing the temperature of the molecules trapped in a MOT or launching the captured molecules in a molecular fountain\cite{Tarbutt2013}, a substantial improvement of the eEDM measurement sensitivity could be realized.

The past few decades have witnessed a tremendous progress in the techniques of deceleration of molecules, such as the direct laser cooling and Stark decelerator\cite{Doyle2004,Carr2009,Jin2012,DeMille2017,Moses2016}. Molecules that are amenable to direct laser cooling are basically diatomics with highly diagonal Franck-Condon factors (FCFs), which would minimize the number of required repumping lasers to form a quasi-cycling transitions\cite{DiRosa2004}. Recently, radiation pressure beam slowing using white-light or chirped techniques has been successfully applied to a few species, such as SrF\cite{Barry2012}, CaF\cite{Zhelyazkova2014,Hemmerling2016}, YO\cite{Yeo2015}, and the 2D or 3D MOT of the three molecules have been realized\cite{Hummon2013,Barry2014,Truppe2017,Anderegg2017}. Reviewing these achievements, we would notice that the capture efficiency of molecules from a beam into a MOT could be limited by the rather long slowing distance because of the loss to higher vibrational dark states and the transverse divergence during the deceleration process. Radiation pressure force that depends on the scattering rate can be significantly reduced by the multiple sublevels and the Doppler shifts and hence leads to the long slowing distance. Apparently, such situation would be even worse for the heavy molecules used for the precise measurements. Therefore, a deceleration mechanism that is different from the radiation pressure one is required to address these troubles. To date, some approaches that extend the traditional Stark decelerator have been proposed or demonstrated to decelerate the heavy polar molecules, such as traveling-wave deceleration\cite{Bulleid2012,VANDENBERG2014}, alternate gradient (AG) focusing methods\cite{Tarbutt2004}, a novel electrostatic Stark decelerator\cite{Wang2016}.

Different from the deceleration method above, the idea of stimulated slowing may also open the door for an efficient and short-distance deceleration of the heavy species. The concept of stimulated slowing has been under extensive investigation since the earliest proposals and experimental demonstrations of the rectified dipole force imposed on atoms in 1980s\cite{Kazantsev1987}. The bichormatic force (BCF), which arises from the rapid coherent sequence of absorption and stimulated emission cycles, was then theoretically developed and experimentally applied to the manipulation of several atom species\cite{Grimm1990,Corder2015}. Since 2011, E. E. Eyler \emph{et al} actively promoted this field forward on both theoretical and experimental frontiers, and made great contributions to describe and examine the performances of BCF on molecules with more complicated structures and internal degeneracies\cite{Chieda2011,Galica2013,Aldridge2016}, from the phenomenological analysis based on two-level models\cite{Chieda2011} to the quantitative treatment for multilevel systems by means of the direct numerical solution for time-dependent density matrix\cite{Aldridge2016}. By numerical simulations, they showed that a cryogenic buffer-gas-cooled beam of CaF could be decelerated nearly to rest without a repumping laser and within a longitudinal distance of $\sim$1 cm. The similar work with MgF\cite{Dai2015,Yang2017} also verified the effectiveness of stimulated slowing of molecular beams. More recently, I. Kozyryev \emph{et al} experimentally demonstrated the transverse deflection of a cryogenic buffer-gas beam of SrOH using the BCF\cite{Kozyryev2018}, which should really indicate the potentials of BCF to decelerate the diatomic and polyatomic molecules for precision measurements or for trapping.

Here we theoretically demonstrate the stimulated longitudinal slowing of heavy-atom molecules by the coherent optical bichromatic force, mainly focusing on the rapid and short-distance deceleration method by keeping a relatively large force exerted on the molecules during the slowing process. On one hand, we will discuss the advantages in magnitude and velocity range of decelerating heavy molecule using BCF. On the other hand, we will illustrate the necessity for maintaining a constant phase-difference between the two couter-propagating beat notes that forms the bichromatic field as the molecules are slowing down. The molecules can be phase-stably slowed under a constant beat phase, which is similar to the concept of longitudinal deceleration with a synchronous phase angle in the electrostatic Stark decelerator\cite{van_de_Meerakker2008}. Taking the YbF molecule that is under exploration for the eEDM measurements by E. A. Hinds group\cite{Hudson2011,Lim2017} as an example, we will show that a cryogenic buffer-gas beam of YbF with a forward velocity of 120 m/s can be decelerated below 10 m/s within a distance of 3.5 cm and with a laser irradiance for each traveling wave of 107.2 W/cm$^2$. An optional experimental design for producing the bichromatic field and maintaining a large BCF magnitude by a phase-difference compensation method is proposed. The elimination of Zeeman dark states and the need for a repuming laser to avoid premature loss of the molecules to noncycling vibrational states are also discussed. Last but not least, a simple approach to estimate the performance of the BCF on some heavy molecules is presented. These molecules are also promising candidates for precision measurements, but some essential molecular constants concerning hyperfine levels are yet to be determined and thus an accurate numerical simulation based on multilevel systems seems difficult. This estimation should be helpful to preliminarily evaluate the possibility and effectiveness of the BCF deceleration of the chosen molecules.

The paper is organized as follows. Sec. \ref{section2} gives a brief introduction to the BCF deceleration of heavy molecules with a phase-difference compensation method, based on which one optional experimental setup are proposed. Sec. \ref{section3} discusses some details for treating the Zeeman or vibrational dark states and shows the simulation results of the bichromatic slowing. Sec. \ref{section4} presents the simple model to estimate the performance of the BCF exerted on some candidate molecules which lack some specific constants for an accurate calculation. Sec. \ref{section5} gives a conclusion to this work.

\section{\label{section2}THEORY AND METHOD}

Description of the bichromatic force with a $\pi$ pulse model has already provided us with an intuitive picture of the interaction mechanism between a bichromatic field (produced by a balanced counter-propagating pair of two-color CW laser beams) and the two-level atoms with a resonance frequency of $\omega$\cite{Soding1997}. In a nutshell, each two-color beam with frequencies $\omega\pm\delta$ can be regarded as a beat note train which can become a series of $\pi$ pulses if the laser power is properly tuned, and a large net force can be produced by alternating cycles of absorption and stimulated emission if the counter-propagating beat notes are sequenced at a proper relative phase. Theoretical study indicates the magnitude of the force for a two-level system can reach $\hbar k\delta/\pi$ with a optimal relative phase of beat notes of $\pi/2$, much larger than the radiation pressure force which gives a maximum of $\hbar k\Gamma/2$ for $\delta\gg\Gamma$. The velocity range of BCF of $\delta/k$ is also more favorable than radiative force that would be significantly reduced by the changing Doppler shift during the deceleration process.

However, in a realistic multilevel system of molecule, there are a few imperfections deriving primarily from the internal splitting, degeneracy and distant dark states, which lead to the complexities in both theoretical study and experimental conditions. While the dark states can be addressed by techniques of destabilization\cite{Berkeland2002} or repumping, the variations in transition strength and frequency shifts from level to level would also cause the reduction of the optimal magnitude of the force, that is, not all of the transitions among the multilevels can be simultaneously driven at the optimal rate with the same laser power and detuning. As there are no analytical solutions to describe this process, a quantitative treatment in multilevel systems by direct numerical solution for the time-dependent density matrix in the rotating-wave approximation proposed by L. Aldridge \emph{et al} should offer a satisfactory insight towards the performance of the BCF on the molecules\cite{Aldridge2016}. We would also adopt this treatment in our theoretical analysis of stimulated slowing of heavy molecules if necessary, although a relative simple but less accurate approach for estimation would also be given in Sec. \ref{section4}.

Another major reduction of the magnitude of the BCF can be caused by the variation of the relative phase of the counter-propagating beat notes as the molecule situates different positions during the deceleration process. Let us reconsider the expression of the bichromatic field aligned along the $z$-axis formed by a balanced counter-propagating pair of two-frequency laser beams with equal electric-field magnitude $E_0$ in all four components. The two components in each beam are symmetrically detuned from a carrier frequency $\omega$ by $\delta$, which actually produce two trains of beam notes travelling in opposite directions:
\begin{align}\label{eq1}
  E(z,t) &=E_-+E_+ \notag\\
         &=2E_0\rm{cos}(kz+\omega t)\rm{cos}(\delta t+\varphi /4)\notag\\
         &+2E_0\rm{cos}(kz-\omega t)\rm{cos}(\delta t-\varphi /4),
\end{align}
where $E_-$, $E_+$ denote the electric fields of left- and right-travelling beat-note train, respectively, $k$ is the wave vector, and $\varphi=4\delta z/c$ is the phase difference between the counter-propagating beat note envelops. In the approximation that the slowing length is much smaller than the beat length, that is, $z\ll c/\delta$, the phase difference $\varphi$ remains approximately constant, as in the cases for CaF, MgF, which are light and amenable to a very short-distance deceleration within $\sim$1 cm\cite{Aldridge2016,Yang2017}. However, this assumption is invalid for heavy-atom molecules with a longer deceleration distance for bichromatic longitudinal slowing, though the distance is still much shorter than that for radiation pressure slowing, as we will see below.

	Now we will take YbF molecule as an example to illustrate the variation of the phase difference $\varphi$ in the deceleration process. This molecule has already been extensively studied and well employed for the eEDM measurements\cite{Tarbutt2013,Hudson2011}, and more recently, the one-dimensional sub-Doppler laser cooling of a YbF molecule beam to 100 $\mu$K has been demonstrated\cite{Lim2017}. With the rotationally-closed main transition $X^2\Sigma_{1/2}^+(v=0,N=1) \leftrightarrow A^2\Pi_{1/2}(v'=0,J'=1/2)$ at 552 nm with a natural linewidth of $\Gamma = 2\pi\times5.7$ MHz\cite{Tarbutt2013}, we calculated the direct numerical solution of the time-dependent density matrix in the rotating-wave approximation for a multilevel system with a lower-state manifold of 12 levels (Zeeman sublevels of the hyperfine structures characterized by $F=2,\ 1^+,\ 1^-,\ 0$) and upper-state manifold of 4 levels ($F=1,\ 0$). The relative electric dipole transition matrix elements $r_{ij}$ between states $\vert i\rangle$ and $\vert j\rangle$ can be conveniently calculated following the method in Ref. \cite{Wall2008} with the $J$ mixing. Although a lager magnitude or a wider velocity range of the BCF should theoretically be obtained with a larger bichromatic detuning $\delta$, we should consider the cost of a laser power requirement that scales as $\delta^2$. Here we set $\delta=50\Gamma$, and it corresponds to an irradiance of each laser beam of 107.2 W/cm$^2$ when the system is driven by an optimal total Rabi frequency amplitude $\Omega^{tot}=\sqrt{3/2}\delta$, whose definition and verification have been given in prior publication\cite{Aldridge2016}. The total power of the bichromatic field produced by four laser beams with a diameter of 1 mm would be about 3.4 W to satisfy the irradiance requirement, which is accessible with the modern laser technologies such as the dye laser at this wave band. In order to destabilize the Zeeman dark states of $m_F=\pm2$ within $F=2$ of the ground states for a $\pi$-polarized laser beam, a skewed static magnetic field with a magnitude of 20 Gauss and an angle of $60^{\circ}$ is utilized in our calculation of the BCF. The detuning of the carrier frequency $\omega$ of the four laser beams from the transition center of mass of the system was set to be $-40$ MHz.

\begin{figure}[]
\includegraphics[width= 0.5\textwidth]{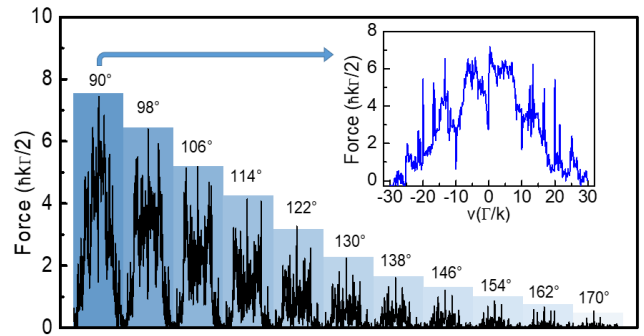}
\caption{\label{figure1}(Color online) The calculated relationships between the BCF exerted on YbF molecules and molecular velocities under different phase differences ($\varphi$) of the counter-propagating beat-note envelops from 90 to 170 degrees. The insert shows the optimal BCF effects with $\varphi=\pi/2$. All the numerical solutions are calculated based on the time-dependent density matrix for 16 levels, with a 20 Gauss magnetic field with a skewed angle of 60$^{\circ}$.}
\end{figure}

The numerically calculated results for the BCF exerted on YbF as a function of molecular longitudinal velocity was presented in Fig.\ref{figure1}. The inset of Fig.\ref{figure1} shows the optimal BCF effects with the optimal phase-difference of the counter-propagating beat note envelops of $\varphi=\pi/2$. In this optimized condition, we can see from the inset that the velocity range of the BCF can still be approximated by $\Delta v\approx\delta/k=50\Gamma/k$, where $\Gamma/k=$3.15 m/s for YbF. The average magnitude of BCF for molecules with near-zero velocities is about $5.6\hbar k\Gamma/2$. Again, the velocity range and magnitude of BCF should indicate the superiority of stimulated slowing over the radiative one. However, Fig.\ref{figure1} also shows how the velocity-dependent BCF varies under different phase-differences from 90 to 170 degrees, corresponding to a molecular travelling distance of $\sim$5 cm for YbF. Obviously the force magnitude would decrease sharply as the molecular beam travels such a short distance, which motivates us to the way of compensating the variation of this phase-difference to maintain a large BCF magnitude while the heavy molecules are being decelerated.

As is well known, in Stark or optical decelerators, the phase stability is essential for the polar molecules to be decelerated under a synchronous phase angle\cite{van_de_Meerakker2008}. Time-varying fields synchronized with the moving molecules are employed to phase-stably decelerate the molecular beams, and the slowing process in the decelerator can be viewed as slicing a bunch of molecules with a narrow spatial distribution and velocity distribution out of the incident molecular beam, and decelerating them to arbitrarily low velocities\cite{Bethlem1999,Bochinski2003}. The analogous concept can be introduced into the stimulated slowing of heavy-atom molecules that requires a relatively long deceleration distance. The molecules can be slowed under a constant beat phase-difference by means of an optical phase compensation method to keep a favorable force over the deceleration process.

\begin{figure}[]
\includegraphics[width= 0.5\textwidth]{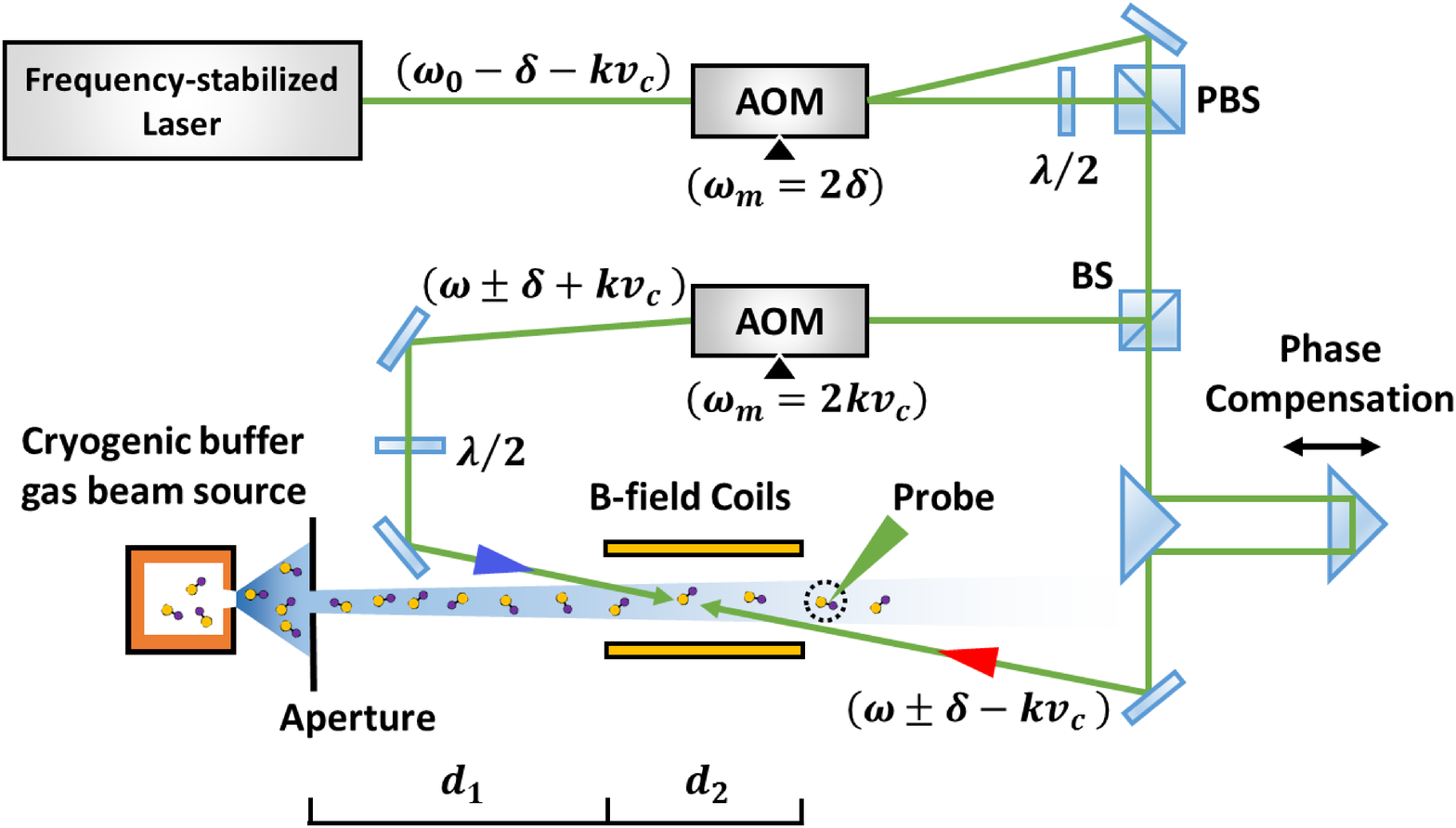}
\caption{\label{figure2}(Color online) Proposed experimental design for producing Dopper-shifted bichromatic field, compensating the phase-difference changes, and producing cryogenic molecular beam with a buffer gas beam source. The expressions labeled within parentheses indicate the laser frequencies after stabilization by some frequency-stabilization method or modulation by the AOMs.}
\end{figure}

According to the discussion above and the practicality of producing both the bichromatic field and a precooled molecular beam, one possible experimental design is proposed in Fig.\ref{figure2}. The molecular beam can be a cryogenic buffer gas beam emitted from a 4 K-helium-filled cell, inside which the YbF molecules are generated and precooled. After being well collimated with a small aperture away from the cell, the beam could eventually have a forward velocity of 100-200 m/s and an RMS transverse velocity spread $\Delta v^{RMS}_t<2$ m/s\cite{Shuman2010}. The molecules will enter the bichromatic slowing region with a length of $d_2$ after passing through a free flight region with a length of $d_1$, and are then probed with some necessary detection techniques following the deceleration. A DC magnetic field with an optimized magnitude and skewed angle relative to the laser polarization is applied throughout the slowing region. The bichromatic field originates from a single-mode laser with a stabilized frequency tuned at $\omega-\delta-kv_c$, where the Doppler frequency shift of $kv_c/2\pi$ is added to center the coverage of BCF at $v_c$ to take full advantage of its wide velocity range. An acousto-optic modulators (AOM) operating at $2\delta$ is employed to produce two frequencies at $\omega\pm\delta-kv_c$, which are combined with a PBS and then split by a BS. One beam passes through another AOM operating at $2kv_c$ to have a frequency of $\omega\pm\delta+kv_c$ and become a blue-tuned right-travelling beat note train, the other beam will serve as a red-tuned one after an optical delay line to adjust the relative beat phase\cite{Soding1997}. Here the optical delay line is also the key for compensating the variation of the beat phase difference during the deceleration process. By synchronizing the position of the delay line with the motion of molecules within a relative short distance, a constant beat phase difference and thus a large BCF can be maintained to effectively decelerate the molecules. The delay line could be translated synchronously by a motorized linear translation stage driven by stepper motors, which is common in modern optical experiments. The required translation length of the delay line is moderate, which is only half of the molecular slowing length.

\section{\label{section3}SIMULATION OF BICHROMATIC SLOWING}

\begin{figure}[]
\includegraphics[width= 0.5\textwidth]{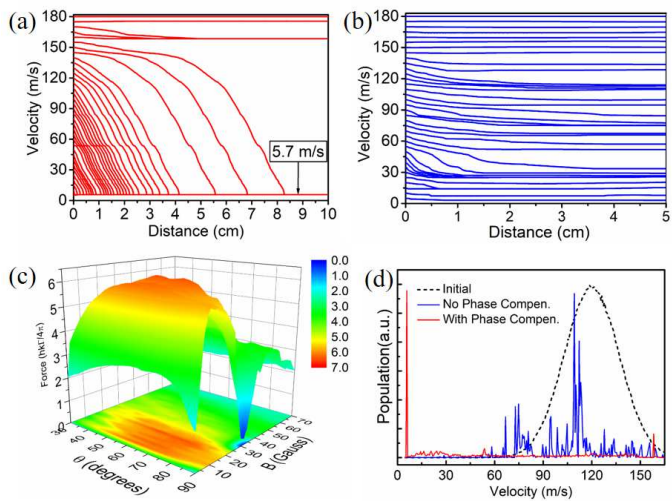}
\caption{\label{figure3}(Color online) (a) The changes of the velocities as a function of slowing distance for a constant beat phase-difference within the slowing distance. The center of the coverage of BCF is set at $v_c=27\Gamma/k=85$ m/s. (b) Velocities as a function of distance for a varying phase without phase compensation. (c) The dependence of the average BCF near zero-velocities on magnitude ($B$) and angle ($\theta$) of a magnetic field to remix the dark states for a $\pi$-polarized laser filed. (d) The initial and final longitudinal velocity distributions with and without compensating the beat phase-difference. The Monte Carlo simulation results are obtained with the optimal parameters discussed in the text and the conditions of the proposed experimental design in Fig. 2. }
\end{figure}

In this section, we will compare the bichromatic deceleration with and without compensating the varying relative beat phase using Monte Carlo simulations. Before that, the treatments for the dark states within transition manifold or the distant incoherently coupled dark states should be considered, which would otherwise cause the BCF to diminish rapidly to zero. It is well known that the Zeeman dark states will always exist in a system where the ground-state multiplicity of degenerate projection quantum states $m$ exceeds that of excited-state by at least two, regardless of the polarization of the laser that couples the two manifolds\cite{Berkeland2002}.  Solutions to this problem have been discussed in Ref. \cite{Aldridge2016,Yang2017} for the BCF in multilevel systems, in which two common schemes including a skewed magnetic field and a rapid polarization switching are investigated. Here for the bichromatic deceleration of YbF, we will adopt the former scheme for the multiple transitions driven by a $\pi$-polarized laser filed, by applying a magnetic field with a magnitude $B$ and an angle $\theta$ relative to the light polarization to remix the dark states. Using the optimal laser irradiance discussed in Sec. \ref{section2} determined from the optimal total Rabi frequency amplitude $\Omega^{tot}$ for $\delta=50\Gamma$, the magnitude and angle of the magnetic field was surveyed with the numerical calculation for a sixteen-level system in order to find the optimum parameters. The dependence of the average BCF near zero-velocities on $B$ and $\theta$ is shown in Fig.\ref{figure3}(c), which indicates an optimal $B = 20$ Gauss and $\theta = 60^{\circ}$. With these parameters, the resultant velocity-dependent force profile has been shown in the inset of Fig.\ref{figure1}.

Whether to use a repumping laser to avoid premature loss of the molecules to distant vibrational dark states has also been discussed in previous publications\cite{Chieda2011,Aldridge2016,Yang2017}, but what we want to emphasize here is that for a heavy molecule to be decelerated by BCF, there is almost no doubt that the repumping laser should be added over the deceleration process. Although molecules with favorable $A-X\ (0-0)$ Frank-Condon factors $f_{00}$ which is typically larger than 0.9 are widely utilized for the laser slowing, they will be lost out of the system in hundreds or even dozens of spontaneous decay cycles without repumping lasers. The interaction time between the molecule and the laser field will be limited by $t_{lim}=\left[P_e\Gamma\left(1-f_{00}\right)\right]^{-1}$ where $P_e$ is the average excited-state population. A reasonable estimation of $P_e\approx1/7$ was verified in early works\cite{Aldridge2016,Yang2017} and also confirmed in our simulation for YbF by averaging the calculated upper-state fraction. Consequently we can estimate the  $t_{lim}\approx2.7\ \mu$s for YbF with $f_{00}=0.928$\cite{Tarbutt2013}. Obviously it is impossible to effectively decelerate a heavy molecule within a moderate distance in such a short time, as can be seen in the following discussion.

To investigate the performance of the BCF slowing of YbF under the optimal parameters for $\delta=50\Gamma$ discussed above including laser irradiance, magnetic field and repumping light, we first surveyed the deceleration effectiveness for molecules with different initial velocities using the velocity-dependent force illustrated in Fig.\ref{figure1}. Assuming that all the molecules begin to be decelerated from the same position, the changes of the velocities as a function of slowing distance are simulated. With the center of the coverage of BCF at the velocity of $v_c=27\Gamma/k=85$ m/s, the simulation results are shown in Fig.\ref{figure3}(a) for a constant beat phase-difference within the slowing distance and Fig.\ref{figure3}(b) for a varying one without a phase compensation method. We can see from Fig.\ref{figure3}(a) that molecules with initial velocities below 140 m/s could be decelerated to the same final velocity of $\sim$5.7 m/s within a distance of 5 cm. Without the phase compensation, however, a 5-cm-long distance corresponding to a relative beat phase change of $\sim80^{\circ}$ would lead to the reduction of BCF that has been shown in Fig.\ref{figure1}. Consequently, the deceleration of molecules yields a poor performance as shown in Fig.\ref{figure3}(b). We also carried out a Monte Carlo simulation with the optimal parameters used above and the conditions of the proposed experimental design given in Sec. \ref{section2}. A buffer-gas precooled and collimated molecular beam with a forward velocity of $v_f=(120\pm20)$ m/s flies freely for a distance of $d_1=5$ cm before being illuminated by the bichromatic field for a distance of $d_2$. The initial and final longitudinal velocity distributions with and without compensating beat phase-difference are shown in Fig.\ref{figure3}(d). A considerable fraction of the YbF molecules can be slowed to a velocity of $\sim5.7$ m/s with a deceleration efficiency of about 1.2\% after passing through a BCF region with $d_2 = 3.5$ cm. Here we set $d_2 = 3.5$ cm because there is a compromise between the transverse divergence and the initial velocities that can be decelerated, since Fig.\ref{figure3}(a) indicates that a longer distance is required to decelerate a molecule with a larger speed. The average time for the deceleration is $\sim1.51$ ms, much larger than the out-of-system decay time $t_{lim}$, so the rempumping lasers for addressing vibrational dark states are indispensable. As discussed in Sec. \ref{section2}, the distance that the optical delay line should be changed to compensate the varying beat phase-difference is just $d_2/2=1.75$ cm. Deceleration result for a non-phase-compensating case is also given in Fig.\ref{figure3}(d), where there are almost no molecules that can be slowed below 10 m/s, as can be expected according to the discussion above.

	For further verification of the effectiveness of bichromatic slowing for heavy molecules with a constant relative beat phase, we conducted similar numerical calculations and simulations on BaF molecule, which has also been regarded as a promising candidate for eEDM measurements\cite{Kozlov1997}. Considering the attainable laser power, we employed the identical laser irradiance as used in YbF's case in the simulation of stimulated slowing of BaF, which corresponds to a bichromatic detuning of $\delta=139\Gamma$ for this molecule. The simulation results and the relevant parameters used in the calculations are summarized in Table \ref{table1}, in which we also list the counterparts of YbF for comparison. It can be inferred from the table that the heavy molecules with a smaller saturation irradiance, $I_s=\pi hc\Gamma/(3\lambda^3)$, and a larger ratio of $\Gamma/k$ will yield larger magnitude and velocity range of BCF, and thus can be decelerated more rapidly and efficiently from a molecular beam with a higher initial forward velocity.

\begin{table}[]
\caption{\label{table1}
Parameters and results in Monte Carlo simulations of bichromatic slowing with a phase-difference compensation (P.C.) method, assuming that the molecular beams to be decelerated have an initial $v_f=(120\pm20)$ m/s and a free-flight distance of $d_1=5$ cm before being slowed. The lifetime of $A^2\Pi_{1/2}$, the Frank-Condon factor and the transition wavelength of $A$-$X$ (0-0) of BaF molecule are from Ref. \cite{Tao2016}. }
\begin{ruledtabular}
\begin{tabular}{lcc}
\ & YbF & BaF\\
\hline
Mass (amu) & 192.1 & 156.3\\
Lifetime of $A^2\Pi_{1/2}$ (ns) & 28 & 56\\
Frank-Condon Factor of $A$-$X$ (0-0) & 0.928 & 0.9508\\
Wavelength of $A$-$X$ (0-0) (nm) & 552 & 860 \\
Saturation irradiance $I_s$ (mW/cm$^2$) & 4.42 & 0.58\\
Irradiance per beam (W/cm$^2$) & 107.2 & 107.2\\
Magnetic field magnitude (G) & 20 & 10\\
Magnetic field angle & 60$^{\circ}$ & 70$^{\circ}$ \\
Force near zero-velocity ($\hbar k\Gamma/2$) & 5.6 & 9.6\\
Velocity range (m/s) & 158 & 339\\
Center velocity for Doppler shift (m/s) & 85 & 73\\
Deceleration distance with P.C. (cm) & 3.5 & 3.0\\
Average deceleration time with P.C. (ms) & 1.51 & 0.564\\
Deceleration Efficiency with P.C. (\%) & 1.2 & 4.1\\
\end{tabular}
\end{ruledtabular}
\end{table}

\section{\label{section4}SIMPLIFIED ESTIMATION OF BCF}

In addition to the molecules mentioned above, there are also other heavy-atom candidates with potentials in the precision measurements, for example, RaF\cite{Isaev2010}, HgF\cite{Prasannaa2015} etc., which have the similar molecular structures to the molecules above and should also amenable to the efficient, short-distance bichromatic slowing with a constant beat phase-difference. However, due to the complexity of the observation and measurement of these promising cold molecules, some essential molecular constants concerning hyperfine levels are yet to be determined and thus an accurate numerical simulation based on multilevel systems seems inaccessible. Therefore, it would be quite worthwhile to find a simplified model to estimate the effects of BCF exerted on these molecules for making a rapid evaluation on the feasibility of experiment. In 2011, E. E. Eyler and co-workers presented a statistical approach to estimate the magnitude of BCF by introducing a weighted degeneracy factor and a so-called ``force reduction factor'' $\eta$ to the expression of BCF derived from the two-level approximation\cite{Chieda2011}, in consideration of the degeneracies of the ground and excited states and the different transition strengths among these levels. In their subsequent theoretical work in 2016, they pointed out that this treatment is much less accurate compared to the multilevel models for CaF, and that there is a substantial overall uncertainty with $\eta$\cite{Aldridge2016}.

\begin{figure}[]
\includegraphics[width= 0.5\textwidth]{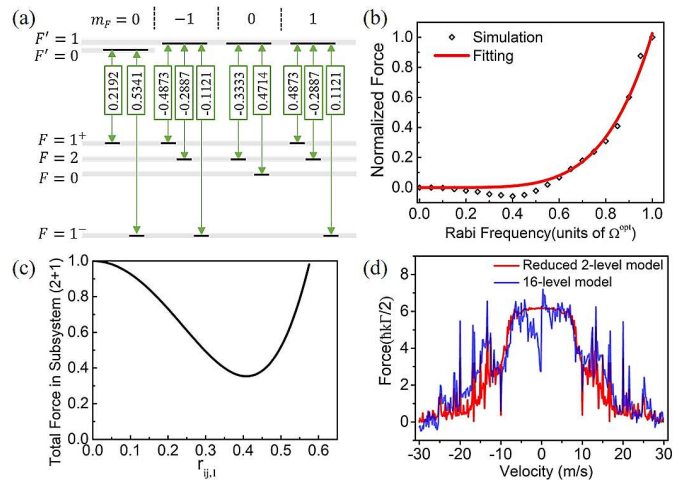}
\caption{\label{figure4}(Color online) (a) The Rabi oscillations of the $X^2\Sigma^+ \leftrightarrow A^2 \Pi_{1/2}\ (0,0)\ Q_{12} (0.5)/P_{11} (1.5)$ branch in YbF for a $\pi$-polarized light. Relative dipole transition matrix elements $r_{ij}$ are labeled for each transition. The entire system can be viewed as two (2+1) and two (3+1) $\Lambda$-type subsystems. (b) The relationship between the BCF and the Rabi frequency (in units of the optimal one) by simulation and fitting. (c) The typical dependence of the total force in a (2+1) subsystem on one of the two transition matrix elements within this subsystem. (d) The velocity-dependent force profiles calculated with the 16-level-based method (blue curve) and with a simple two-level model but reduced by the 0.228 factor (red curve). }
\end{figure}

Here we will present a simple approach to estimate the performance of the BCF exerted on some candidate molecules which possess the similar level structures to YbF or CaF but lack some specific constants for a full-level simulation. We envision that the diatomic molecules discussed here consist of a zero spin isotope of the heavy atom and a $^{19}\rm{F}$ anion with a nuclear spin of $I=1/2$\cite{Sauer1996,Prasannaa2015}, and the angular momenta couplings of these molecules are described by the same Hund's cases as those of YbF or CaF molecules. The same main transition between $X^2\Sigma^+$ and $A^2\Pi_{1/2}$ states is also employed for the stimulated slowing.

Considering the primary factor that reduces the magnitude of BCF, we will see that in a realistic multilevel system, it is impossible for every component of the transitions to obtain a $\pi$ pulse even if an optimal power is employed, due to the different transition line strength from level to level. The BCF is an average over the effects of several concurrently cycling transitions with distinct strengths, which depend on the different electric dipole coupling between the lower and upper states, as can be seen from Fig.\ref{figure4}(a), where the Rabi oscillations of the $X^2\Sigma^+ \leftrightarrow A^2 \Pi_{1/2}\ (0,0)\ Q_{12} (0.5)/P_{11} (1.5)$ branch of YbF for a linearly polarized light are sketched. We can conclude from Fig.\ref{figure4}(a) that the same laser field will simultaneously drive multiple transitions at different Rabi frequencies. However, things can be simpler for a model of four weighted subsystems as shown in Fig.\ref{figure4}(a). In this model, by isolating the $\Delta m_F=0$ transitions for a $\pi$-polarized light, the entire system can be viewed as two (2+1) and two (3+1) $\Lambda$-type subsystems which are weakly coupled to each other by incoherent radiative decays\cite{Aldridge2016,Yang2017}. The quadrature sum of the $r_{ij}$ within each of the four subsystems is $\sum_ir^2_{ij}=1/3$, which is inherently common for $\pi$-polarized-light driven molecules with the same level structures as YbF or CaF, so the same laser irradiance can oscillate all four subsystems at the same optimal total Rabi frequency amplitude $\Omega^{tot}=\sqrt{3/2}\delta$. However, the two or three transitions within each of the four subsystems will certainly not work at the optimal Rabi frequency, because the Rabi frequency amplitude of each single transition $\Omega^0_{ij}$ within each subsystem is determined by
\begin{align}\label{eq2}
  \Omega^0_{ij}=\frac{r_{ij}}{\sqrt{\sum\limits_ir_{ij}^2}}\Omega^{tot},
\end{align}
where $j$ represents the single upper state in each of the four sublevels, and $i$ the lower two or three states. Since $\sum_ir^2_{ij}=1/3$ and thus the absolute value of $r_{ij}$ cannot exceed $1/\sqrt{3}$, each single transition merely oscillates at a Rabi frequency that is a fraction of the optimal $\Omega^{tot}$, and hence the magnitude of BCF is reduced compared to a two-level system. To investigate the force reduction due to the deviation from the optimal Rabi frequency, we calculated the relationship between the force magnitude and the Rabi frequency based on the two-level model, and then fitted it with a curve, as shown in Fig.\ref{figure4}(b). We can see that the force will be half of the maximum if working at a Rabi frequency just ~13\% deviated from the optimal one.

From Eq. \ref{eq2} we can see that the Rabi frequency at which one transition can work is determined by its dipole transition matrix element $r_{ij}$, so the total force in each of the four subsystems can be predicted with the values of $r_{ij}$ of the two or three transitions within that subsystem. Fig.\ref{figure4}(c) shows the dependence of the total force in a (2+1) subsystem on one of the two  dipole transition matrix elements, noting that if one matrix element is given, the other will be determined since  $\sum_ir^2_{ij}=1/3$. Averaging the force in the (2+1) system gives $F_{(2+1)}=0.33F_{2\textrm{-}level}$, where $F_{2\textrm{-}level}$ is the magnitude of force obtained with two-level model. A similar procedure can be implemented on the (3+1) subsystem, which will give $F_{(3+1)}=0.16F_{2\textrm{-}level}$. The results from each subsystem were then combined with a weighting given by the fraction of ground-state sublevels included in each subsystem, yielding an average BCF magnitude that is about 22.8\% of that derived from the two-level model. To verify the validity of this estimation for a reduced BCF with respect to the ideal two-level case, we drew the velocity-dependent force profile calculated with the 16-level-based method (the same as the inset of Fig.\ref{figure1}) and the one calculated with a simple two-level model but reduced by a 0.228 factor at the same frame in Fig.\ref{figure4}(d), where we can see a favorable match between them both in magnitudes and in velocity ranges. Furthermore, we used this simple approach to estimate the force profile of CaF and MgF, whose molecular constants have been well determined, and then compared the results with the one given by the numerical solution for the time-dependent density matrix based on 16-levels. As a consequence, we also find satisfactory matches between them. Therefore, we believe that the simplified approach of the estimation of the BCF presented here can be extended to the molecules with the same level structures as them, such as RaF, HgF. Without having to know some specific molecular hyperfine constants, it is possible to make a rapid evaluation on the experimental feasibility of the bichromatic slowing of the candidates.

\section{\label{section5}CONCLUSION}

We have theoretically demonstrated the possibility of the stimulated longitudinal slowing of heavy-atom molecules by the coherent optical bichromatic force. Several highlights are summarized as below.

(1) Taking the YbF meolecule as an example, we show the advantages in magnitude and velocity range of decelerating heavy molecule using BCF. Using an bichromatic detuning of $\delta=50\Gamma$ and an optimized dark-state-eliminating magnetic field of 20 Gauss and $60^{\circ}$, the velocity range of the BCF can be 158 m/s and the average magnitude of BCF with near-zero velocities can be $5.6\hbar k\Gamma/2$.

(2) We explain the necessity and method for maintaining a constant phase-difference between the two couter-propagating beat notes as the molecules are slowing down, and an optional experimental design is proposed accordingly.

(3) We show that a rapid and short-distance deceleration of heavy molecules by phase-compensation method can be realized. A beam of YbF with a forward velocity of 120 m/s can be decelerated below 10 m/s within a distance of 3.5 cm and with a laser irradiance of 107.2 W/cm$^2$, which is attainable with modern lasers. By comparing the simulation results of YbF with those of BaF, we conclude that the molecules with a smaller saturation irradiance and a larger ratio of $\Gamma/k$ are amenable to a more rapid and efficient stimulated slowing.

(4) A simple approach to estimate the performances of the BCF on some heavy molecules is presented. The BCF exerted on molecules which possess the similar level structures to YbF or CaF but lack some specific constants for a full-level simulation can be obtained by a two-level model reduced by a simple factor that is caused by the deviation from working at the optical Rabi frequency. This estimation should be helpful to preliminarily evaluate the possibility and effectiveness of the BCF deceleration of the candidate molecules.

In conclusion, the stimulated longitudinal slowing of heavy-atom molecules could be a promising approach to break through the space constraint or the limited capture efficiency of molecules loadable into a MOT in traditional molecular deceleration schemes. The prospects of decelerating a large fraction of molecules to the capture velocity of a 3D MOT will pave the way for a significant improvement of the precision measurement sensitivity by further reducing the temperature of the molecules or launching them in a molecular fountain\cite{Tarbutt2013}.

\begin{acknowledgments}

We acknowledge support from the National Natural Science Foundation of China under grants 91536218, 11374100, the Natural Science Foundation of Shanghai Municipality under grant 17ZR1443000, and the outstanding doctoral dissertation cultivation plan of action of ECNU under Grant No. PY2015026.

\end{acknowledgments}

\bibliography{stimulated_slowing}

\end{document}